\documentstyle[preprint,revtex]{aps}

\begin{document}
\thispagestyle{empty}


\begin{center}
\hfill{IP-ASTP-10-94}\\
\hfill{June 1994}\\

\vspace{1 cm}

\begin{title}
Gravitational Wave Induced Large-scale Polarization of \\Cosmic Microwave
Background Radiation
\end{title}
\vspace{1 cm}

\author{Ka Lok Ng and Kin-Wang Ng}
\vspace{0.5 cm}

\begin{instit}
Institute of Physics, Academia Sinica\\ Taipei, Taiwan 115, R.O.C.
\end{instit}
\end{center}
\vspace{0.5 cm}

\begin{abstract}
We discuss the contribution of gravitational wave to the cosmic microwave
background radiation (CMBR) anisotropy and polarization.
It is found that the large-scale polarization of CMBR is less than 1\% for a
standard recombination universe.
The effect of matter reionization will enhance the CMBR polarization to a 10\%
level.  We have computed the CMBR polarization for two
extreme cases (not absolutely ruled out) and found that
further enhancement of the ratio is possible.
We conclude that measuring
the polarization of CMBR on large-angular scales can
probe the ionization history of the early universe, set constraints on
baryon density and the spectral index of the gravitational waves.

\vspace{0.5 cm}
\noindent
{\it Subject headings: cosmology: cosmic microwave background --- cosmology:
observations}
\end{abstract}
\newpage


The large-scale anisotropy of the cosmic microwave background radiation (CMBR)
measured by the DMR onboard the Cosmic Background Explorer satellite (Smoot
et al. 1992) may be induced by
the density perturbations (scalar mode) and the primordial gravitational waves
(tensor mode) via the
Sachs-Wolfe (SW) effect (Sachs \& Wolfe 1967).
Recently, it was argued that the anisotropy might
be dominated by the tensor mode (Krauss \& White 1992).  Furthermore, it
was shown that the
tensor-mode dominance actually occurs in certain inflation models
(Davis et al. 1992).
It was suggested that by comparing large- and
small-scale anisotropy measurements, one can separate the scalar- and
tensor-mode contributions (Davis et al. 1992; Crittenden et al. 1993).

The polarization of CMBR is another clue that could have a great
potential of probing the early universe.  There have been several works on
calculating
the small-angular scales ($\le 1^o$) r.m.s. polarization of CMBR induced by
adiabatic density perturbation, assuming a standard recombination universe
(Kaiser 1983, Bond \& Efstathiou 1984), and in a re-ionized universe
(Bond \& Efstathiou 1984, Nasel'skii \& Polnarev 1987).  It was shown that
roughly 10-20$\%$ of CMBR anisotropy is polarized on arc-minute scales.

The calculation for both large- and small-
angular scale polarization of CMBR induced by adiabatic and isocurvature
density perturbations has been performed (Bond \& Efstathiou 1987).
An analytic estimation of the quadrupole polarization induced by scalar and
tensor metric perturbations was made in various cosmological models including
early matter reionization (Ng \& Ng 1993).
Similarly, the large-scale polarization of CMBR induced by the scalar and
tensor modes within inflationary models was investigated
under the assumption of no reionization (Harari \& Zaldarriaga 1993).
The r.m.s. temperature
anisotropy and polarization of the CMBR induced by the tensor mode
perturbation of arbitrary wavelength was also computed (Frewin et al. 1993).
A detailed numerical calculation of scalar and tensor contribution to the
CMBR polarization power spectrum in inflationary models
has also been carried out (Crittenden, Davis, \& Steinhardt 1993).
It was concluded that the polarization can reach a 10 $\%$ level and may be a
useful discriminant for determining the ioniza
In this paper we will give a detailed numerical calculation of the large-angle
polarization to anisotropy ratio of CMBR induced by long-wavelength
tensor modes in an universe with and without a late re-ionization era.
We consider the large angular effect of tensor mode only since the absolute
magnitude of polarization on small-angular scales is negligibly small although
the degree of polarization may be large. It is
because sub-horizon gravitational waves at decoupling time which contribute to
small-scale CMBR fluctuations disperse and redshift away.
We solve numerically the equation of motion for the gravitational
wave and use it as an input in the relativistic collisional Boltzmann equation
for the photon.
Here we will not stick to a specific inflation model, rather assuming a generic
power spectrum of gravitational waves. Our tensor mode result can be directly
compared to the scalar mode contribution.

We shall use the units $c=\hbar=1$ throughout.
The metric that we use is of the flat Robertson-Walker form

\begin{equation}
ds^2=a^2(\eta) \left(d\eta^2-d{\bf x}^2 \right) \;,
\end{equation}
where $a(\eta)$ and $d\eta=dt/a(t)$ are the scale factor and conformal
time respectively.
Here we normalize the conformal time to unity today.
In this metric, $\Omega_{\rm total}=\Omega_{CDM}+\Omega_B=1$,
where $\Omega_{CDM}$ and $\Omega_B$ denote respectively the cold dark and
baryonic matter. By solving the Friedmann equation, we obtain

\begin{equation}
\eta=\left[\left(1+\frac{1+z_{eq}}{1+z}\right)^{1\over 2}-1\right]
     \left[(2+z_{eq})^{1\over 2}-1\right]^{-1},
\end{equation}
where $z_{eq}$ is the redshift when the matter and radiation densities are
equal.

To study how polarized photons propagate in the universe, one need to
solve the equation of transfer for photons (Chandrasekhar 1960).  In general,
arbitrarily polari${\bf n}=(n_l,n_r,n_u,n_v)$, where $n=n_l+n_r$ is the
distribution
function for photons with $l$ and $r$ denoting two directions
at right angle to each other.

The equation of transfer for an arbitrarily polarized photon
is governed by the collisional Boltzmann equation,

\begin{eqnarray}
\left({\partial \over \partial \eta}+{\bf e}\cdot {\partial \over \partial
\bf x} \right) {\bf n}
 &=& - {1 \over 2}{ {\partial{\bf n}} \over {\partial{\rm ln} \nu} }
              { {\partial h_{ij} }\over{\partial \eta} }e^i e^j  \nonumber\\
 &-&\sigma_T N_e a
      \left[ {\bf n} - {1 \over {4 \pi}} \int_{-1}^1 \int_0^{2\pi}
      {P(\mu,\phi,\mu^{'},\phi^{'}) {\bf n} d\mu^{'} d\phi^{'}} \right] \;,
\end{eqnarray}
\label{n}
where $\sigma_T$ is the Thomson scattering cross section, $N_e$ is
the number of free electrons per unit volume,
($\mu={\rm cos} \theta, \phi$) are the polar angles
of the propagation direction ${\bf e}$ of the photon with a comoving frequency
$\nu$,
and $P$ is the phase matrix for Thomson scattering.

We consider the anisotropy and polarization
of CMBR due to tensor perturbation with a power spectrum
$A_{\bf k}^2 \propto k^{n_t}$,
where $\bf k$ is the wave vector of the tensor mode
and $n_t=0$ corresponds to a scale invariant spectrum.
For the tensor perturbation we have
$h^{\lambda}_{ij}=\int d{\bf k} h e^{i{\bf
k}\cdot{\bf x}}\epsilon^{\lambda}_{ij}$, where
$\epsilon^{\lambda}_{ij}$  denote the gravitational wave polarization
tensors,
$\epsilon^{+}_{ij}= \epsilon_i\epsilon_j-\epsilon^{*}_i\epsilon^{*}_j$, and
$\epsilon^{\times}_{ij}= \epsilon_i\epsilon_j^*+\epsilon^{*}_i\epsilon_j$;
where $\epsilon_i$ and $\epsilon_i^*$ are two mutually orthogonal unit vectors
perpendicular to the wave vector $\bf k$.
The gravitational wave amplitude we employ in our calculation then satisfies,

\be{d^2h \over d\eta'^2} + {4(\eta'+1) \over \eta'(\eta'+2)} {dh \over d\eta'}
                      + k'^2 h = 0
\end{equation}
\label{h}
where
$\eta' \equiv (\sqrt {2} -1)\eta / \eta_{eq}$,
$k' \equiv k \eta_{eq}/ (\sqrt{2} -1)$, and $\eta_{eq}$
is the radiation-matter equality time.

The solution ${\bf n}$ for the equation of transfer assumes the form
${\bf n} ={\bf n_0} + {n_0 \over 2}\delta {\bf n}$,
where ${\bf n_0}$ and $\delta {\bf n}$ are the unperturbed solution and
perturbation respectively. We expand $\delta {\bf n}=\int d{\bf k} {\bf n'}
e^{i{\bf k \cdot x}}$, where ${\bf n'} = \alpha {\bf a} + \beta{\bf b}$,
${\bf a}= {1\over 2}(1-\mu^2){\rm cos}2\phi (1,\;1,\;0)$, and
${\bf b}= {1 \over 2} ( (1+\mu^2){\rm cos}2\phi,\;
                        -(1+\mu^2){\rm cos}2\phi,\; 4\mu {\rm sin}2\phi )$
for the $\lambda=+$ mode solution, with ${\rm cos}2\phi \leftrightarrow {\rm
sin} 2\phi$ for the $\lambda=\times$ mode.

Substituting the solution ${\bf n}$ into equation~$(\ref{n})$, we obtain two
coupled differential equations
for $\xi=\alpha+\beta$ and $\beta$, namely,
\begin{equation}
\dot{\xi}+[q+ik\mu]\xi = H
\end{equation} \label{xib}
\begin{equation}
\dot{\beta} + [q + ik\mu]\beta  = F,
\end{equation} \label{be}
\noindent
where $q=\sigma_T N_e a$,
$
F(\eta) = \frac{3}{16}q \int_{-1}^{+1}
\left[ \left(1+\mu'^{2}\right)^{2}\beta - \frac{1}{2}\xi\left(1-\mu'^{2}
\right)^{2} \right]d\mu',
$
and
$
H(\eta) = dh/d\eta.
$
We solve the two coupled differential equations,
equations~$(\ref{xib})$ and~$(\ref{be})$ by expanding
$\xi$ and $\beta$ in terms of Legendre polynomials, i.e.,
\begin{eqnarray}
\xi(\mu') = \sum _{l} (2l+1) \xi_l P_l(\mu'),  \\
\beta(\mu')=\sum _{l} (2l+1) \beta_l P_l(\mu').
\end{eqnarray}

Hence, we obtain a system of coupled differential equations,
\begin{eqnarray}
& &{ d\xi_0 \over d\eta }= -q\xi_0 - ik\xi_1 + & &{ d\beta_0 \over d\eta }=
-{3\over 10}q\beta_0 - ik\beta_1
     + q ( {5 \over 7 } \beta_2 + {3 \over 35} \beta_4 -{1 \over 10} \xi_0
          +{1 \over 7} \xi_2 - {3 \over 70} \xi_4)  \\
&&{\rm for} \;l \; \ge 1,      \nonumber \\
& &{ d\xi_l \over d\eta }= -q \xi_l - {ik \over 2l+1 }
                           (l \xi_{l-1} + (l+1)\xi_{l+1}) \\
& &{ d\beta_l \over d\eta }= -q \beta_l - {ik \over 2l+1 }
                           (l \beta_{l-1} + (l+1)\beta_{l+1}) \;,
\end{eqnarray}
where the term $dh/d\eta$ is input from solving numerically
equation~$(\ref{h})$.

To describe the degrees of anisotropy and polarization, we will
compute the power spectra for the anisotropy, $C^{\alpha\alpha}_l$,
and polarization, $C^{\beta\beta}_l$.
To evaluate these, we expand the
photon fluctutation distribution function in terms of spherical harmonic
functions, i.e., $\delta{\bf n} = \sum_{l,m} {\bf a}_{lm} Y_{lm}$, hence
${\bf a}_{lm} = \int \delta{\bf n} Y^*_{lm} d\Omega $.

The total power spectrum is given by $\langle \sum_m {\bf a}_{lm}^\dagger {\bf
a}_{lm}\rangle = C^{\alpha\alpha}_{l} + C^{\beta\beta}_{l}$, where
$\langle\rangle$ denotes the sky average.
After a lengthy calculation, we obtain
the anisotropy power specrtum (Crittenden et al. 1993),

\begin{eqnarray}
C_l^{\alpha\alpha}
&=&\pi \int d{\bf k} k^{n_t-3} (2l+1)(l-1)l(l+1)(l+2)
\nonumber \\
& &\mid { {A_{l-2}^*} \over {(2l-1)(2l+1)} }
       - { {2A_l^*} \over {(2l-1)(2l+3)} }
       + {  { A_{l+2}^*} \over {(2l+1)(2l+3)} } \mid ^2 \;, \nonumber \\
\end{eqnarray}
where $A_i=\alpha_{i+} + i\alpha_{i\times}$.
Similarly, the polarization power spec
\begin{eqnarray}
\lefteqn{ C_l^{\beta\beta}= {8\pi (2l+1) \over (l-1)l(l+1)(l+2)} }\nonumber\\
&& \int d{\bf k} k^{n_t-3} {1\over 2}\left| \left[
\sum^{ \left[{l-2 \over 2}\right]}_{m=0}{{(l-2m-3)(l-2m-2)} \over
(2l-4m-5)}B_{l-2m-4}
\right. \right.                                             \nonumber        \\
&+& {  {6[(l-2m-1)(l-2m-2)-{2 \over 3}](2l-4m-3)}  \over {(2l-4m-5)(2l-4m-1)} }
B_{l-2m-2}   \nonumber \\
&+& \left. { {(l-2m-1)(l-2m)} \over {(2l-4m-1)} } B_{l-2m} \right] \nonumber
\\
&-&\left.{ {l(l-1)} \over 2} \left[{ {(l-1)l} \over {(2l-1)(2l+1)} } B_{l-2}
   + { 2(3l^2+3l-2) \over {(2l-1)(2l+3)}} B_l
+{{(l+1)(l+2)} \over {(2l+1)(2l+3)}} B_{l+2}\right] \right|^2 \nonumber \\
&+&\left| \sum^{\left[{l-2 \over 2}\right]}_{m=0}
  2 \left[ (l-2m-2) B_{l-2m-3} +(l-2m-1) B_{l-2m-1}\right] \right.
\nonumber \\
&-&\left.{ { (l-1)l} \over (2l+1) } \left[ lB_{l-1} + (l+1)B_{l+1}
\right] \right|^2 \;,
\end{eqnarray}
where $B_i = \beta_{i+} + i\beta_{i\times}$, [$l$-2/2] signifies the integral
part of $l$-2/2, and the summation is cut
off at the first term with a negative subscript.  It is interesting to notice
that the cross term, $C^{\alpha\beta}_l$, vanishes.
The values of the multipole moments $\alpha_i$ and $\beta_i$ are obtained by
solving numerically the system of differential equations, equations (9) to
(12).

We compute the
ratio of polarized
to unpolarized moments, $ C_l^{\beta\beta} / C_l^{\alpha\alpha} $
for $l$ from 2 to 200 for two cases: instantaneous hyderogen recombination
and late matter re-ionizaion. Higher moments are neglected because they are
relatively small.  To ensure
we have captured the dominant contribution to the power spectra, $k$
covers the range from 0.1 to 5$l$.
In standard recombination, the polarization is less than 1\%. However, it
will be greatly enhanced by re-ionization.
The universe is hPeterson 1965,
Partridge 1980). The energy required for this reionization could be supplied by
the radiation of young galaxies (Ozernoi \& Chernomordik 1976, Tegmark \& Silk
1994).
Also, re-ionization can smooth out excessive temperature fluctuations on degree
and sub-degree angular scales predicted in standard cold dark matter model in
order to reconcile the model with all observational CMBR limits (Sugiyama, Silk
\& Vittorio 1993).

Fig. 1 shows the normalized power spectrum for the
anisotropy.  The Hubble constant is $H_0=100 \;h\;{\rm km\;s^{-1}\;Mpc^{-1}}$
with $h=0.5$ and the baryon density is given by $\Omega_Bh^2=0.0125$.
The constant behavior of $l(l+1)C_l$ for small $l$ with a scale invariant
spectrum ($n_t=0$) in standard recombination ($Zh=0$) is evident
from the figure.  A consequence of reionization ($Zh=30-90$ and $n_t=0$) is
that
the amplitude of the temperature anisotropy on small scale will be reduced.
For a universe with a high baryon density (e.g. $\Omega_B=0.125$),
the damping on anisotropy by reionization is prominant.  We also studied
the non-scale invariant case with $n_t=1$, which is marginal
consistent with the COBE two years result (Bennett et. al 1994). In
inflationary models, $n_t$ is related to the power index of energy density
perturbation $n_s$ by $n_s=n_t+1$.
Our calculation indicates that the anisotropy due to a spectrum with a high
power index combined with reionization can actually imitate the result in
standard recombination.

Fig. 2 shows the polarization to anisotropy ratio as a function of $l$ and
reionization.  The polarization to anisotropy ratio increases quite
signifiant for an early time reionization history, but not sensitive to the
$n_t$ value.  We notice that the anisotropy to polarization ratio for $l=2$
is of the order $(1\sim 3)10^{-5}$ for $our earlier analytic calculations for
the tensor mode contribution to the
$\beta \over \alpha$ ratio (Ng \& Ng 1993), which is approximately
$(3 \sim 5)10^{-5}$.
Note that the polarization to anisotropy ratio peaks around $l\sim 10-40$
which correspond
to a few degree angular scales. For a high baryon density universe,
the ratio is further increased, however, the
absolute magnitude of polarization is still small since the anisotropy drops
even drastic.

Actual observations measure the correlation function, which is defined as,
$C(\Theta) = {1 \over {4\pi}}\sum_l C_l W_l P_l(\cos \Theta)$, where $W_l$
is the window function, and $\Theta$ is the separation angle.
In measurement the lower end of $l$ is excluded by limited sky coverage,
whearas the high-$l$ cutoff is fixed by the finite beam width.
In Table 1, we list the r.m.s. total polarization-to-anisotropy ratio
$[C^{\beta\beta}(0) / C^{\alpha\alpha}(0)]^{1/2}$ from $l=5$ to $50$ with
$W_l=1$, which should correspond to typical
large-angular-scale measurement.  The r.m.s ratio has no discernible
difference when $l$ is summed to 200.

In conclusion, we have calculated the contribution of gravitational wave
to the CMBR, for both scale invariant and non-scale invariant cases,
as well as the effects of matter re-ionization and baryon density.
The results are shown in Figs. 1 and 2.  We found that the polarization to
anisotropy ratio is largely enhanced as a result of matter re-ionization,
and it reaches a 10\% level.  A high baryon density universe and a non-scale
invariant spectrum of gravitaitional wave could further increase this ratio.
Currently, precise polarimeter is being built, with 100 times more sensitivity
than previous polarization experiments, in an attempt to measure
the polarization (P. Timbie, private communication).
Hence, large-angular-scale pomeasurable.  Any trace of polarization or a better
limit will prove
invaluable to probe the initial condition of the universe.

\vspace{2 cm}
\noindent{{\bf Acknowledgements}}
\vspace{0.5 cm}

This work was supported in part by the R.O.C. NSC Grant No.
NSC83-0208-M-001-053 and NSC82-0208-M-001-131-T.

\noindent
\begin{table}
\caption{Polarization to Anisotropy Ratio}

\begin{tabular}{ccc}
$Zh $&$\Omega_B$&$[{{C^{\beta\beta}(0)} \over {C^{\alpha\alpha}}(0) }]^{1/2}$
\\
\hline
$0$&$0.05 $ &$0.0025 $\\
$30$&$0.05 $ &$0.022 $\\
$60$&$0.05 $ &$0.058 $\\
$90$&$0.05 $ &$0.090 $\\
$90$&$0.125 $ &$0.13 $\\
$90$&$0.05 (n_t=1)$ &$0.11 $\\
\end{tabular}
\label{table1}
\end{table}

\noindent
{\bf \Large References}

\noindent
Bennett, C. L., et al. 1994, (preprint) [LANL bulletin board:astro-ph/9401012]

\noindent
Bond, J. R., \& Efstathiou, G. 1984, ApJ, 285, L45

\noindent
Bond, J. R., \& Efstathiou, G. 1987, MNRAS, 226, 655

\noindent
Chandrasekhar, S. 1960, Radiative Transfer (New York: Dover)

\noindent
Crittenden, R., et al. 1993, Phys. Rev. Lett., 71, 324

\noindent
Crittenden, R., Davis, R. L., \& Steinhardt P. J. 1993, ApJ, 417, L13

\noindent
Davis, R. L., et al. 1992, Phys. Rev. Lett., 69, 1856

\noindent
Frewin, R., Polnarev, A., \& Coles, P. 1994, MNRAS, 266,
L21

\noindent
Gunn, J. E., \& Peterson, B. A. 1965, ApJ, 142, 1633

\noindent
Harari, D., \& Zaldarriaga, M. 1993, Phys. Lett. B, 319, 96

\noindent
Kaiser, N. 1983, MNRAS, 202, 1169

\noindent
Krauss, L., \& White, M. 1992, Phys. Rev. Lett., 69, 869

\noindent
Nasel'skii, P., \& Polnarev, A. 1987, Astrofizika, 26, 543

\noindent
Ng, K. L., \& Ng, K.-W. 1993, IP-ASTP-08-93 (preprint) [LANL bulletin board:
astro-ph/9305001]

\noindent
Ozernoi, L. M., \& Chernomordik, V. V. 1976, Astr. Zh. 53, 459 [Sov. Astron.,
20, 260]

\noindent
Partri
\noindent
Polnarev, A. G. 1985, Astron. Zh., 62, 1041 [Sov. Astron., 29(6), 607]

\noindent
Rees, M. J. 1968, ApJ, 153, L1

\noindent
Sachs, R. K., \& Wolfe, A. M. 1967, ApJ, 147, 73

\noindent
Smoot, G., et al. 1992, ApJ, 396, L1

\noindent
Sugiyama, N., Silk, J., \& Vittorio, N. 1993, ApJ, 419, L1

\noindent
Tegmark, M., \& Silk, J. 1994, CfPA-94-th-24 (preprint) [LANL bulletin board:
astro-ph/9405042]

\noindent
White, M. 1992, Phys. Rev. D, 46, 4198

\newpage

\noindent
{\bf Figures Caption}

Fig.1  Normalized anisotropy multipoles as a function of $l$.  The four
solid curves corrspond to the reionization at $Zh=0, \; 30, \;60, \; 90$
with $\Omega_B=0.05$.  The dotted and dash curves denote the cases where
$Zh=90, n_t=1$, and $Zh=90, \Omega_B=0.125$ respectively.

Fig.2  Ratio of polarization moments to anisotropy moments as a function of
$l$.  The four solid curves corrspond to the reionization at
$Zh=0, \; 30, \;60, \; 90$
with $\Omega_B=0.05$.  The dotted and dash curves denote the cases where
$Zh=90, n_t=1$, and $Zh=90, \Omega_B=0.125$ respectively.

\end{document}